\documentclass[aps,prl, twocolumn]{revtex4}
\usepackage{graphics}

\begin{document}

\title{Hysteresis loops of Co-Pt perpendicular magnetic multilayers}
\author{David L. Mobley}
\affiliation{Department of Physics, University of California, Davis, CA 95616}
\author{Christopher R. Pike}
\affiliation{Department of Physics, University of California, Davis, CA 95616}
\author{Joseph E. Davies}
\affiliation{Department of Physics, University of California, Davis, CA 95616}
\author{Daniel L. Cox}
\affiliation{Department of Physics, University of California, Davis, CA 95616}
\author{Rajiv R. P. Singh}
\affiliation{Department of Physics, University of California, Davis, CA 95616}

\date{\today}

\begin{abstract}
We develop a phenomenological model to study magnetic hysteresis in two samples designed as possible perpendicular recording media. A stochastic cellular automata model captures cooperative behavior in the nucleation of magnetic domains. We show how this simple model turns broad hysteresis loops into loops with sharp drops like those observed in these samples, and explains their unusual features. We also present, and experimentally verify, predictions of this model, and suggest how insights from this model may apply more generally.
\end{abstract}

\pacs{75.60.Ej, 75.70.Ak, 75.50.Ss, 75.60.Jk}
\keywords{hysteresis, nucleation, Cobalt-Platinum, cellular automata, model, reversal curves}

\maketitle

Magnetic thin films with magnetization oriented perpendicular to the plane of the film have become attractive as magnetic bits get smaller and smaller. Cobalt-platinum multilayers are thin films with several desirable characteristics for future recording media, such as large ratio of coercive field (field where $M=0$) to saturation magnetization and sharp initial drops in their hysteresis loops. Although grain sizes are too large at present for applications, they are nevertheless interesting model systems, both as high coercivity thin films~\cite{kn:Weller}, and as a testing ground for Ising model based simulations of hysteresis in real materials.

\begin{figure}
\includegraphics{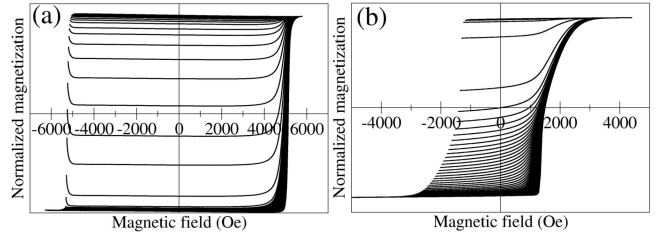}
\caption{Major hysteresis loops (envelope) and first order reversal curves for two samples provided by B. Terris at IBM. Sample (a) shows sudden sharp drop in major hysteresis loop, while sample (b) shows a similar drop as the magnetization begins to fall, but then has a broad tail.}
\label{fig:forcs}
\end{figure}

These films are also of fundamental interest as interacting dynamical systems. In particular, the precise onset of magnetization reversal, which has been observed before in Co-Pt multilayers~\cite{kn:Weller, kn:Phillips, kn:Hatwar, kn:Bennett, kn:DellaTorre}, suggests the shape of the hysteresis loop may be dominated by nucleation. This is the case for some magnetic thin films of interest in recording~\cite{kn:Phillips, kn:Hatwar, kn:Mansuripur}. Nucleation phenomena are of broad interest, e.g. in phase transitions~\cite{kn:Langer}, materials synthesis~\cite{kn:Fan}, and protein aggregation~\cite{kn:Slepoy}. Unresolved issues for these Co-Pt multilayers include the origin of the broad tail in the hysteresis loop near magnetic saturation and the drop in first-order reversal curves following reversal of the magnetic field (Fig.\ \ref{fig:forcs}). Additionally, differences between nominally similar samples are not well understood.

Here we obtain the major hysteresis loop and first-order reversal curves (measured by decreasing the magnetic field along the major hysteresis loop until a predefined reversal field, then tracing out a reversal curve as the field is again increased to saturation) for two Co-Pt multilayer samples, then develop a lattice-based theoretical model to explain key features of these measurements. The model makes testable predictions concerning time-dependence of the reversal curves, which we test, and temperature-dependence of the hysteresis loop. Our model is remarkable not so much in terms of quantitative agreement with experiment, but in the physical insight it provides into a variety of experimental results.

Two Co-Pt multilayer samples were provided by B. Terris at IBM. These samples are similar in composition to those described by Weller et al.~\cite{kn:Weller} and consist of a 20 nm Pt buffer layer, followed by 10 Co(0.6 nm)/Pt(1 nm) bilayers. Growth temperatures for the multilayers were 157$^\circ$C and 320$^\circ$C for the samples of Fig.\ \ref{fig:forcs}(a) and (b), respectively. Reversal curves (Fig.\ \ref{fig:forcs}) were taken with an Alternating Gradient Magnetometer (Princeton Measurements Corporation). The general procedure for each reversal curve was to decrease the magnetic field every 0.1s until reaching a predefined reversal field. Data was then collected at steps along the reversal curve with an averaging time at each data point of 0.13s. Variations on this procedure were used to test predictions of our model and will be discussed below.
 
The experimental hysteresis loops and reversal curves for these samples exhibit several unusual features. To our knowledge, a simple explanation of these has not yet been provided. Particularly, we note the broad tail (as the magnetization approaches saturation) shown in Fig.\ \ref{fig:forcs}(b). Modeling in terms of exchange coupling can produce sharp drops due to an energy barrier associated with nucleation of reverse domains, but the rounding at the tail is typically comparable to the rounding at the onset (as the magnetization begins to fall)~\cite{kn:Miyashita, kn:Zhu}. We notice here that the tail is much broader than the onset. This has been observed previously~\cite{kn:Phillips, kn:Bennett, kn:DellaTorre} and described as an asymmetry around the coercive field~\cite{kn:DellaTorre}. Additionally, we notice a drop in the reversal curves in Fig.\ \ref{fig:forcs}(a) following reversal of the magnetic field. That is, the magnetization continues to fall even after the magnetic field begins to increase. Reversal curves are less commonly examined but this has been observed in an unrelated material~\cite{kn:Basso} and without comment in a similar material~\cite{kn:DellaTorre}. These features will be explained here. 

Sethna et al. have argued that one should be able to trade realistic microscopic degrees of freedom for a set of rules as part of a renormalization procedure to study different classes of hysteretic behavior at long length and time scales~\cite{kn:Sethna}. While some multiparameter micromagnetic (time-dependent Landau-Lifshitz-Gilbert equation) models can produce hysteresis loops similar to those observed here~\cite{kn:Lyberatos}, we here present a simpler model which explains these features in a simple, physically transparent manner and, we believe, captures much of the essential physics of these samples. 

Stochastic cellular automata models, like we use here, have been used successfully before to model nucleation barriers~\cite{kn:Slepoy}. Ours involves spins -- representing individual magnetic bits that are either up or down -- on a 2D triangular lattice, as these materials are designed to work for 2D storage. The key model ingredients are as follows: 

1.\ \emph{Anisotropy field:} Each spin is initially assigned an anisotropy field, defined as  $H_k=2 K/(\mu_0 M_s)$ where $K$ is the anisotropy energy. The field $-H_k$ can be thought of as the field at which an isolated spin would flip from up to down, and conversely for $+H_k$. Thus each spin has a coercivity and is inherently hysteretic. We assign different $H_k$ values to each spin, typically according to a lognormal distribution. 

2.\ \emph{Energy barrier:} At each step, we first treat each spin independently and compute an energy barrier for it to flip down. For an up-spin, in the case of magnetic field $H>-H_k$, we use $\Delta E=\frac{A H_k}{2} (1+\frac{H}{H_k})^2$, where A is a constant, and for a down-spin in the case of $H<H_k$ there is simply a sign change inside the quadratic~\cite{kn:Bertotti}. The barrier is 0 otherwise. 

3.\ \emph{Attempt probability:} We evaluate the probability of the individual spin attempting to flip, $P=exp(-\Delta E/k_B T)$, where $T$ is an input parameter, and determine whether the spin attempts to flip. We do this for every spin on the lattice and keep a list of those that attempt. 

\begin{figure}
\includegraphics{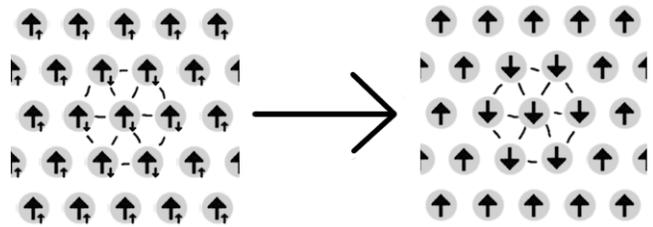}
\caption{Left, a group of seven spins as shown, where each of the seven attempts (small arrows) to flip down in the same step, is the smallest group of spins that can stably flip down (right) in our rules. Thus our rules provide a nucleation barrier to formation of a domain of down spins near M=1, and similarly when near M=-1 for formation of a domain of up spins.}
\label{fig:rules}
\end{figure}

4.\ \emph{Cooperative rules:} We then apply rules which require cooperativity between spins and establish a nucleation barrier. The model we use is as follows: For a spin to flip from up to down, it must have at least three neighbors that are already down or will flip down in the same step (and similarly going from down to up). Thus the smallest size group of spins that can flip down is 7, as in Fig.\ \ref{fig:rules}. This effectively builds in a nucleation barrier and allows us to explore what effect this has on the hysteresis loop. These rules play a diminishing role after initial nucleation.

5.\ \emph{Magnetization and field:} After updating every site on the lattice, we calculate and output the magnetization and step the magnetic field. This represents one simulation step. We then repeat this procedure until reaching saturation.

It is important to note that we do not wait until the sample equilibrates before stepping the field. We thus link a simulation step with some small amount of real time, $\tau$, which has been done successfully before to model kinetic effects in hysteresis loops~\cite{kn:Rao}. The relevant timescale for this may be the ``attempt frequency'' for crossing an energy barrier, which is generally considered to be on the order of a nanosecond~\cite{kn:He}.

\begin{figure}
\includegraphics{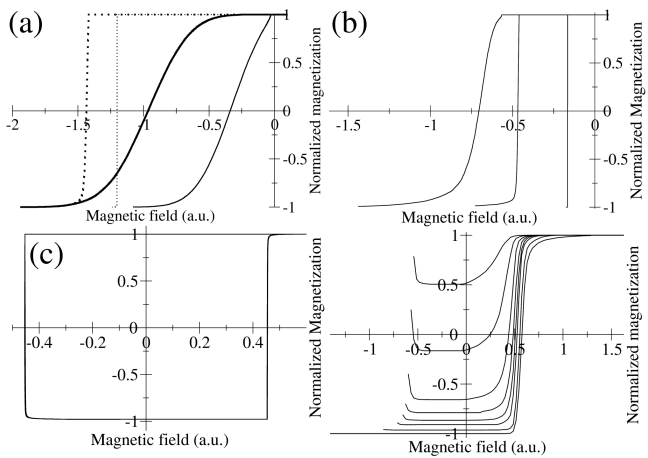}
\caption{(a) Descending half of hysteresis loops with and without cooperative rules for two different temperatures, T=0.04 (thicker lines) and T=0.1 (thinner lines). The portion for increasing H is symmetric. The broad curves (solid lines) are the results when all the spins act independently; the sharp curves (dotted lines) show the effect of the nucleation barrier established by our cooperative rules. (b) A sequence of descending half hysteresis loops at different temperatures (T=0.03, 0.1, and 0.3 from left to right). Notice the overall sharpening at higher T and the much broader tail at lower T. (c) Calculated first-order reversal curves at T=0.1. Shown here are reversals at M=0.9, 0.5 and 0, but the curves essentially all overlap because in all three, even following reversal, magnetization continues to drop nearly to M=-1 before stabilizing. (d) For T=0.05, first-order reversal curves are more reasonable. Note that the magnetization does drop following reversal. We find that the amount of this drop decreases at lower temperatures or for more broad distributions of $H_k$.} 
\label{fig:effect} 
\end{figure}

At the most basic level, we show that without cooperativity, our rules yield hysteresis loops that are quite broad. These loops acquire a sudden drop in magnetization due to the nucleation barrier when we apply the cooperative rules (Fig.\ \ref{fig:effect}(a)). We also find that the temperature dependence of this effect is somewhat unusual (Fig.\ \ref{fig:effect}(b)), in that the drop is actually sharper at higher temperature and both the drop and the tail are broader at low temperature.

The explanation for these effects follows: The sudden drop in the hysteresis loop is due to the relief of the constraints of the cooperative rules following nucleation of domains. That is, once the nucleation barrier is crossed, domains can grow, and it is much easier to add additional spins to a domain (a spin that is still up neighboring a down domain already has some down neighbors and so it is easier for it to have sufficient down neighbors to flip down) than it was to nucleate the domain in the first place. 
 This is similar to a previous description of reversal in terms of two coercivities, a nucleation field $H_n$ necessary to nucleate reversed domains and a field $H_p$ necessary to overcome domain wall pinning. If $H_n>H_p$ reversal can occur very rapidly following nucleation~\cite{kn:Weller, kn:Phillips}. However, our model is different in that we do not have a single pinning coercivity. We refer to the large cascade of spin-flips following nucleation as ``avalanching,'' as described in random-field Ising models~\cite{kn:Avalanches}, although there is a distinction in that in the Ising model, reversed domains tend to pull their neighbors along with them, while here the only driving force is the external field. 

This avalanching applies to the temperature dependence of the drop in that at high temperature, more thermal energy is available to allow the avalanches to proceed. More thermal energy means bigger avalanches and sharper hysteresis loops, because it helps to overcome outlying $H_k$ values that would tend to pin domain walls. This pinning is possible because the rules do provide some constraint even after nucleation. Higher temperature means the disorder is less important and so avalanches are bigger. Low internal disorder has been shown to produce sharp hysteresis loops with large avalanches before~\cite{kn:Avalanches, kn:Berger}, but our model includes both internal disorder and the effects of thermal fluctuations, and shows how temperature affects this sharpness.

The broadening of the tail at low temperatures arises from a freezing in of the hysteresis loop to follow the actual distribution of $H_k$ at low temperatures. That is, in the tail of the hysteresis loop, our rules play little role (most remaining up spins have enough down neighbors they can flip as soon as they ``want'') and spins act independently, so the tail samples the distribution of $H_k$ and it must follow it more closely at low temperatures. This can be seen in Fig.\ \ref{fig:effect} by the rejoining of the T=0.04 curves in the tail region. This tail behavior emerges naturally from our model. It has been observed before and its origin has been hinted at in phenomenological Preisach modeling~\cite{kn:DellaTorre} but we have provided a simple physical explanation.  

We also find (Fig.\ \ref{fig:effect}(c),(d)) that our model does produce a drop in the first-order reversal curves. At reasonably high temperatures (i.e. T=0.1 in our model) it greatly exaggerates this effect: reversing the magnetic field at M=0.9, the magnetization continues to fall almost to M=-1 before flattening out (Fig.\ \ref{fig:effect}(c)). Lower temperatures or broader distributions of $H_k$ make this effect more reasonable (Fig.\ \ref{fig:effect}(c)). Additionally, our model predicts that we can make this effect smaller by increasing the field sweep rate. That is, taking bigger steps in magnetic field gets back to the field necessary to stop the reversal curves from dropping sooner, so they do not fall as far. It is simply due to avalanching -- once a domain nucleates, it can grow easily and it requires getting back to a significantly less negative field to stop the avalanche of domain growth. Consider, for example, a spin with a low $H_k$ which, if isolated, would be easy to flip. If it has several higher $H_k$ neighbors it will not flip until long after it would like to flip. But once its neighbors flip it is quite happy to flip, even if the magnetic field reversal begins, since it is long past the point at which it would have liked to flip on its own. This also means that in our model, this asymmetric tail is present even when the distribution of anisotropy fields is symmetric (i.e. Gaussian).  

\begin{figure}
\includegraphics{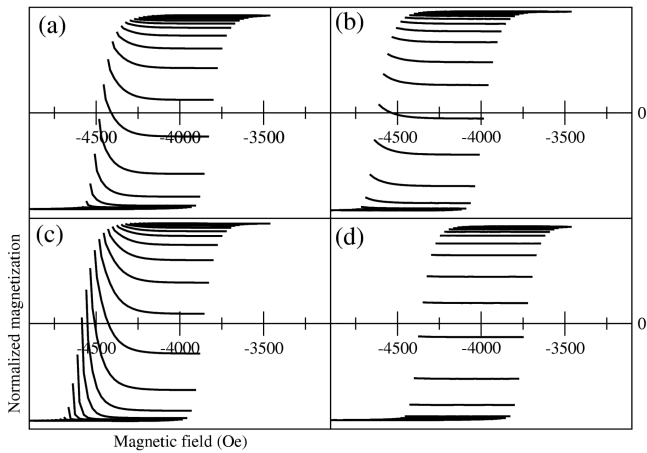}
\caption{Close-ups of the region where the reversal curves drop show the time-dependence of the effect for the sample of Fig.\ref{fig:forcs}(a). Essentially, when there is more pausing before or during measurement, the magnetization drops farther. (a) The normal reversal curves are taken by decreasing the field in 0.1 second steps, then increasing and averaging for 0.13s at each field point. (b) Averaging for 3s at each field point gives avalanches more time to proceed and the magnetization drops further. (c) Equilibrating for 60s prior to taking data on the reversal curve eliminates the effect but puts the magnetization lower at a given field than (a) and (b). (d) As (b) but the field sweep time prior to reversal is also slowed to 3s.}
\label{fig:confirm1}
\end{figure}

With this explanation in mind, experimentally, one should be able to allow the sample to equilibrate before taking a reversal curve and the effect should disappear, since it is a kinetic effect. On the other hand, slowing the field sweep rate along the reversal curve should allow avalanches to proceed further and the effect should get bigger. This was tested experimentally and confirmed for sample (a) (Fig.\ \ref{fig:confirm1}) and the drop also appears for sample (b) if the field sweep rate is slowed enough~\cite{kn:Sample2}. If we assume our theoretical model is capturing this effect accurately, $\tau$ provides the conversion between simulation steps and experimental time. Thus a crude estimate of the effective $\tau$ for these samples is on the order of a few milliseconds. This is obviously much larger than the typical value of around a nanosecond and would suggest that these samples are unusual. This is consistent with the observation that this drop is not typically observed in experimental reversal curves.

\begin{figure}
\includegraphics{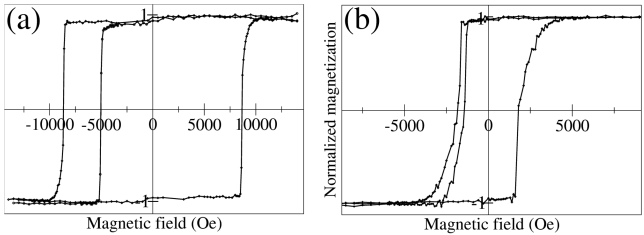}
\caption{Descending half of the major hysteresis loops for our Co-Pt 
multilayer samples as measured at two different temperatures in a SQUID magnetometer. Curves with most negative coercivity are 5 K; the other are at 300K. The 5K curves also show the ascending half of the major loop. (a) The same sample as Fig.\ \ref{fig:forcs}(a) shows a significantly larger coercivity at lower temperature (leftmost curve) and a  broader tail, while sample (b) shows a slightly broader tail (leftmost curve) but little change in coercivity.}
\label{fig:squid}
\end{figure}

To check our prediction of tail broadening at low temperature, we used a Quantum Design Superconducting Quantum Interference Device (SQUID) magnetometer in the standard (DC) operating configuration with the sample perpendicular to the applied field. Our procedure was to measure a full hysteresis loop at 5K and then just the descending half of the loop at 300K, to compare the broadening of the tail. Results are shown in Fig\ \ref{fig:squid}; both samples show a broadening of the tail at lower temperatures in qualitative agreement with the predictions of Fig.\ \ref{fig:effect}(b).  This is more pronounced for sample (a). However, our model's hysteresis loops have a coercivity that is strongly temperature-dependent, but this is true for only one of these two samples. 

Overall, our results are especially significant in that we qualitatively explain the experimentally observed drop in the first-order reversal curves as a kinetic effect due to avalanching and provide experimentally verified suggestions for reducing or increasing this effect. This drop has been observed in another material and attributed to thermal relaxation~\cite{kn:Basso}-- thermal growth of domains at fields below that necessary to induce domain depinning. While thermal energy allows this drop to be larger in our model, we would argue both on theoretical and experimental grounds that our mechanism is distinct. Experimentally, we find that kinetics strongly influence this effect. And in our model, even at T=0, there is still a nucleation barrier which can result in avalanches and produce a drop in reversal curves -- the drop is simply smaller. 

Additionally, our cooperative rules provide hysteresis loops with an asymmetric tail. This asymmetry is due to nucleation, because the nucleation barrier has a profound effect on the initial drop in magnetization. The effects of our cooperative rules disappear by the tail, allowing the tail to still be broad and individual spins to act independently. This effect has only been captured before in vastly more complicated models~\cite{kn:Lyberatos} or by parameter-tuning~\cite{kn:DellaTorre}. 

There are two possible origins of the experimentally observed broad tail. The first is the mechanism we describe here, which does not depend on having an asymmetric distribution of $H_k$. To our knowledge, this effect is absent in models dealing in terms of exchange coupling, as exchange coupling strong enough to produce a nucleation barrier will also tend to prevent a broad tail. However, strongly asymmetric distributions of $H_k$ can produce an asymmetry without hysteresis cooperativity. Independent measurements of the distribution of anisotropy fields will be very useful to distinguish between these two cases.

Our model suggests the importance of inherently hysteretic spins, or hysterons, rather than Ising spins, where hysteresis comes solely from interactions between spins. In samples like these, with strong uniaxial anisotropy, there may be two separate energy scales, one associated with the energy barrier to flip individual grains, and one governing the exchange coupling between neighboring grains which determines a cooperative nucleation barrier. The fact that nucleation is an irreversible process, as described above, helps to create a broad tail in our model despite the absence of any significant rounding at the top, and without invoking asymmetric distributions. Preliminary work suggests that this may remain true in an interacting hysteron model. On the other hand, models based on Ising spins may not capture this behavior, as the deviation from fully polarized state at the top and bottom of the hysteresis loop would be due to reversible flips of outliers and thus would be symmetric. Hence, we suggest that interacting hysteron models, or cooperative nucleation models of the type discussed here, will be better for describing the hysteresis in materials with broad asymmetric tails. 


In conclusion, our model allows us to qualitatively explain important features in the hysteresis loops of these samples and predict kinetic effects observed experimentally. We explain the drop in the reversal curves and the broad tail. Our results suggest that exploring Ising-type models with exchange coupling and inherently hysteretic spins may give new insight into materials like these.

We thank B. Terris for providing our samples. We also gratefully acknowledge discussions with R. Scalettar and G. Zimanyi, and K. Liu for help with the SQUID work. We acknowledge support of the National Science Foundation NEAT-IGERT program (IGERT Grant DGE-9972741) (DLM), the U.S. Department of Energy office of Basic Energy Sciences Division of Materials Research (DLC), and NSF-DMR9986948 (RRPS).

\end{document}